\documentclass[aps,prx,reprint,nofootinbib,preprintnumbers,amsmath,amssymb,amsfonts]{revtex4-2}

\usepackage[utf8]{inputenc}
\usepackage{graphicx}
\usepackage{hyperref}

\newcommand{\Mpl}{M_{\textrm{Pl}}}

\newcommand{\LUV}{\Lambda_{\textrm{UV}}}

\newcommand{\feff}{f_{\textrm{eff}}}
\newcommand{\fEDE}{f_{\textrm{EDE}}}

\begin{document}

\title{Constraints on Early Dark Energy from the Axion Weak Gravity Conjecture}

\author{Tom Rudelius}
\email{rudelius@berkeley.edu}
\affiliation{Physics Department, University of California, Berkeley CA 94720 USA}

\date{\today}

\begin{abstract}
A popular proposal for resolving the Hubble tension involves an early phase of dark energy, driven by an axion field with a periodic potential. In this paper, we argue that these models are tightly constrained by the axion weak gravity conjecture: for typical parameter values, the axion decay constant must satisfy $f < 0.008 \Mpl$, which is smaller than the axion decay constants appearing in the vast majority of early dark energy models to date. We discuss possible ways to evade or loosen this constraint, arguing that its loopholes are small and difficult to thread. This suggests that it may prove challenging to realize early dark energy models in a UV complete theory of quantum gravity.
\end{abstract}

\pacs{}

\maketitle

\section{Introduction}\label{sec:INTRO}

Early dark energy (EDE) was introduced in \cite{Poulin:2018dzj}, and in \cite{Poulin:2018cxd} it was shown to resolve the Hubble tension \cite{Riess:2016jrr, Bonvin:2016crt, Birrer:2018vtm, Planck:2018vyg}. The model features an axion $\chi$ sitting in a potential of the form
\begin{equation}
V(\chi) = \Lambda^4 \left( 1 - \cos \chi/f \right)^n\,.
\label{EDEpotential}
\end{equation}
The axion begins during radiation domination near its maximum at $\chi \approx  \pi f$, and it remains there for a while due to Hubble friction. During this phase, $\chi$ has an equation of state of $w_\chi \approx -1$, and contributes to the vacuum energy (hence the name, early dark energy).

When $H$ decreases sufficiently, the field begins to oscillate around its minimum at $\chi = 0$. For $n=1$, this is the classic axion dark matter scenario: coherent oscillations of the axion act as a source of dark matter, with $w=0$. For more general $n$, one instead has \cite{Turner:1983he}
\begin{equation}
w_n = \frac{n-1}{n+1}\,.
\end{equation}
For $n=2$, for instance, $w_n = 1/3$, and coherent oscillations decay like radiation. Resolving the Hubble tension requires $n \geq 2$. Some typical values for the parameters are \cite{Poulin:2018cxd}:
\begin{align}
f \approx 0.1 \Mpl\,,~~~\Lambda^4 \approx 10^{10} \Mpl^2 H_0^2\,,
\label{eq:typical}
\end{align}
with $\chi_0 \lesssim \pi f$, and $n=2$ or $3$.

In this paper, we will address the question of whether the above model can be embedded into a UV complete theory of quantum gravity. Said differently, we want to know if EDE resides in the landscape or the swampland.

Several aspects of EDE models point in favor of a possible UV completion. Axions are ubiquitous in string compactifications \cite{Svrcek:2006yi, Arvanitaki:2009fg, Baumann:2014nda}, and their connection to the absence of global symmetries in quantum gravity suggests that this is not merely a lamppost effect \cite{Heidenreich:2020pkc}. Axions have periodic potentials, $V(\chi) = V(\chi+ 2 \pi f)$,\footnote{This periodicity condition is modified in axion monodromy models, as the potential develops multiple branches $V_i(\chi)$ satisfying $V_i(\chi + 2 \pi f) = V_{i+1}(\chi)$. For further explanation, see e.g. \cite{Silverstein:2008sg, McAllister:2008hb, Kaloper:2008fb,Kaloper:2011jz}.} which naturally allows them to be much lighter than the Planck mass and protects their potentials from Planck-suppressed higher-dimension operators, such as $\chi^n/\Mpl^{n-4}$.\footnote{Barring significant fine-tuning, these effects will spoil EDE models with scalar fields that are not axions, so even power-law EDE potentials $V(\chi) \propto \chi^{2n}$ \cite{Poulin:2018cxd, Agrawal:2019lmo} must involve axion fields or else violate naturalness.}

On the other hand, some aspects of EDE models point against the possibility of a UV completion. The most glaring issue is the $n \geq 2$ constraint: the potential must be approximately quartic near its minimum (i.e., the axion must be approximately massless). This is hard to arrange for the following reason: axion potentials in UV complete frameworks admit an expansion in higher harmonics of the form
\begin{equation}
V(\chi) = V_0 + \LUV^4 \sum_{k=1}^\infty e^{-S_k} \left( 1 - \cos \left(\frac{\chi }{ k f } + \delta_k \right)  \right) \,.
\end{equation}
Here, $\LUV$ is some UV scale, such as the Planck scale or the string scale, and $\delta_k$ is a phase. In a controlled scenario, $S_1$ is large, and $S_k$ typically grows roughly linearly with $k$, so the dominant contribution to the axion potential comes from the lowest harmonic $k=1$, and all higher harmonics can be ignored. This situation readily occurs, for instance, if the potential $V(\chi)$ is generated by Euclidean D-branes or worldsheet instantons wrapping cycles of a Calabi-Yau manifold in a string compactification, or if the potential is generated by Yang-Mills instantons. This controlled scenario will not give the necessary potential for EDE, however: it yields a potential of the form in \eqref{EDEpotential}, but with $n=1$. 

A potential satisfying the $n \geq 2$ constraint therefore requires a delicate cancelation between distinct harmonics of the potential, such that the $\chi^2$ term of the potential vanishes to very high precision, while the $\chi^4$ or $\chi^6$ terms remain relatively large. Meanwhile, the instanton actions $S_k$ must remain large, or else the UV scale $\LUV$ must be tuned significantly for agreement with \eqref{eq:typical}.  Within string theory, it is unclear how to do all of this while maintaining perturbative control of the instanton expansion.

One way around this issue is to add a second scalar field $\phi$, sometimes referred to as a saxion, which couples to the axion $\chi$ via a potential of the form \cite{Alexander:2019rsc}
\begin{equation}
V(\phi, \chi) = \LUV^4 e^{-(S_0 + \beta \phi)} \left(1 - \cos(\chi/f)\right) + V_0 e^{ \lambda \phi / \Mpl} \,.
\label{axiodilaton}
\end{equation} 
Note that the potential here has $n=1$: it is quadratic near its minimum.
Potentials of this form do show up in string compactifications, with an additional caveat that the axion decay constant $f$ is $\phi$-dependent, $f = f(\phi)$. \cite{Alexander:2019rsc} showed that a potential of this form could produce a cosmologically viable model of early dark energy: an important step towards a UV completion of EDE.

We will see below, however, that even under the most favorable circumstances, there is an additional obstacle to UV completion: the axion weak gravity conjecture. In the following section, we will introduce this conjecture. We will then show that this conjecture leads to a tight, $n$-independent constraint on EDE models, which for typical parameter values amounts to a bound $f < 0.008 \Mpl$ on the axion decay constant. This value is smaller than those typically considered in previous studies of EDE, and the resulting bound is tighter than those previously noted in discussions of swampland constraints on EDE  \cite{Kaloper:2019lpl, Hill:2020osr, McDonough:2021pdg}.
We will elaborate on possible ways to evade this constraint by taking advantage of various ``loopholes'' and uncertainties in the parameter values, and we will conclude with a discussion of directions for future research.


\section{The Axion Weak Gravity Conjecture}\label{sec:WGC}

The axion weak gravity conjecture is part of a larger \emph{swampland program}, which seeks to delineate the boundary between the \emph{landscape} of effective field theories that are compatible with quantum gravity and the \emph{swampland} of effective field theories that are not. This task is nontrivial due to the enormity of the landscape and the sophisticated nature of quantum gravity, and as a result, few conclusions about quantum gravity can be drawn with certainty. Instead, there has been a proliferation of conjectures suggesting that certain features of known quantum gravity theories should be universally present in the landscape.

Not all of these conjectures are on equal footing. Some, such as the absence of global symmetries in quantum gravity \cite{Hawking:1974sw, Zeldovich:1976vq, Zeldovich:1977be}, are supported by very general arguments in string theory (the best candidate for a UV complete theory of quantum gravity) as well as semiclassical arguments (which remain agnostic about the UV completion of quantum gravity). 
Others, such as the de Sitter conjecture \cite{Obied:2018sgi}, are evidently violated in the standard model \cite{Denef:2018etk, Choi:2018rze} and are very difficult to test outside certain controlled regimes of string theory, so their regime of validity may is likely limited to a portion of the landscape.

The axion weak gravity conjecture is somewhere in between. It is a close relative of the weak gravity conjecture \cite{ArkaniHamed:2006dz}, which was originally motivated by black hole physics and is now supported by very general arguments in perturbative string theory \cite{Heidenreich:2016aqi, Montero:2016tif}, many examples in non-perturbative string theory \cite{Lee:2018urn,Lee:2019tst, Alim:2021vhs}, qualitative arguments in effective field theory \cite{Heidenreich:2017sim, Cordova:2022rer}, and several (not entirely convincing) semiclassical arguments \cite{Kats:2006xp, Hod:2017uqc, Crisford:2017gsb, Cheung:2018cwt, Montero:2018fns, Charles:2019qqt, Horowitz:2019eum, Arkani-Hamed:2021ajd}. The axion weak gravity conjecture, in contrast, does not have a direct link to black hole physics, and the primary evidence for it comes from many examples in string theory \cite{Banks:2003sx, Rudelius:2014wla, Bachlechner:2015qja, Conlon:2016aea, long:2016jvd}, plus the lack of a counterexample, and from its close connection to the ordinary weak gravity conjecture \cite{ArkaniHamed:2006dz}.

The precise statement of the axion weak gravity conjecture is also debated \cite{Heidenreich:2015nta, Hebecker:2016dsw, Hebecker:2018ofv}. In four dimensions, the conjecture holds that in a consistent theory of quantum gravity with an axion field with decay constant $f$, there must exist a instanton with action $S$ carrying integer charge $k$ under the axion field such that
\begin{equation}
\frac{f S}{k} \leq c \Mpl \,.
\label{axionWGC}
\end{equation}
Here, the coefficient $c$ is some $O(1)$ number, but its precise value is debated. In theories without exactly massless scalar fields, however, the only finite value of $c$ which has been proposed in the literature is 
\begin{equation}
c = \frac{\sqrt{6} \pi}{4}\,,
\label{ceq}
\end{equation}
which was introduced in \cite{Hebecker:2016dsw, Hebecker:2018ofv} by analogy with black hole physics, and subsequently studied in \cite{Andriolo:2020lul}. For more details, see section 3.7 of \cite{Harlow:2022gzl}.

In what follows, we will therefore define the axion weak gravity conjecture to be equation \eqref{axionWGC} with $c$ given by \eqref{ceq}, though the reader should keep in mind that the precise value of $c$ is a matter of ongoing discussion and research.


\section{Constraints on Early Dark Energy}\label{sec:CONSTRAINTS}

Consider an EDE model with action
\begin{equation}
S= \int d^4x \left[ - \frac{1}{2} (\partial_\mu \chi)^2 - \LUV^4 e^{-S}  \left( 1 - \cos \left( \frac{k \chi}{f}  \right) \right)^n  + ...   \right] \,.
\end{equation}
Here, the instanton action $S=S({\boldsymbol\phi})$ may be a function of one or more scalar fields in the theory, as in \eqref{axiodilaton}. $\LUV$ is some UV scale, such as the Planck scale or the string scale, $k$ is the integer charge of the instanton (a.k.a. instanton number), and $f$ is the axion decay constant.

We define the effective decay constant as $\feff= f/k$ and the axion mass parameter as
\begin{equation}
m_\chi^2 = \frac{\LUV^4 e^{-S} }{\feff^2} \,.
\end{equation}
For $n=1$, $m_\chi$ is precisely the axion mass, whereas for $n >1$, it is a conventionally-defined parameter of the theory with dimension of mass.

\begin{center}
\begin{figure}
\includegraphics[width=70mm]{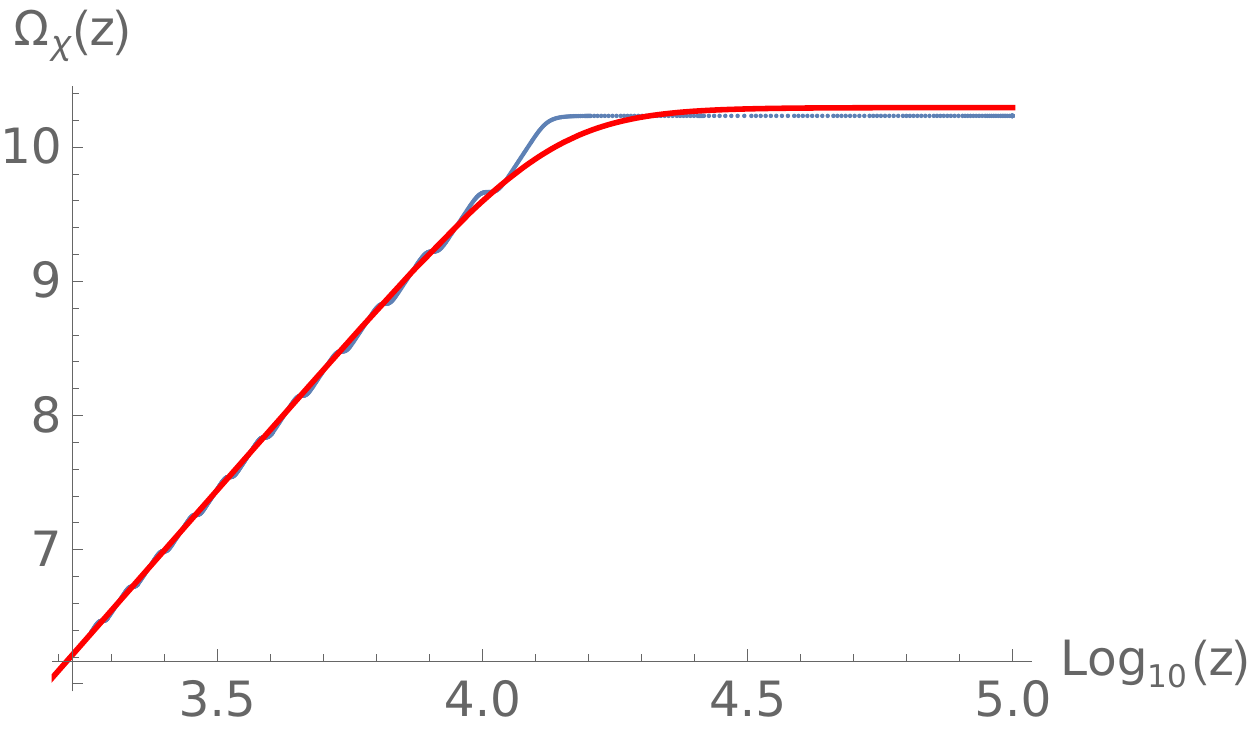}
\caption{Best-fit analysis. In blue, the energy density $\Omega_\chi(z)$ of the axion is plotted as a function of (the log of) the redshift $z$ for a model with $n=3$, $f=0.008 \Mpl$. In red, the results of the best-fit analysis to determine the associated EDE model parameters $\fEDE$, $\Omega_\chi(a_c)$.}
\label{otherfig}
\end{figure}
\end{center}

The axion weak gravity conjecture \eqref{axionWGC} then translates into a bound on the axion mass parameter
\begin{equation}
\feff \log \left( \frac{ \LUV^4}{\feff^2 m_\chi^2} \right)  \leq \frac{\sqrt{6} \pi }{4} \Mpl \,,
\label{generalbound}
\end{equation}
where we have used the value $c = \sqrt{6}\pi/4$ from \eqref{ceq}. Note that this bound is independent of $n$. Plugging in a typical value of $m_\chi \approx 10^{-27}$ eV and setting $\LUV = M_{\textrm{GUT}} \approx 10^{16}$ GeV, we find a bound on the decay constant $\feff$ of 
\begin{equation}
\feff \lesssim 0.008 \Mpl\,,
\label{feffbound}
\end{equation}
which is smaller than the decay constants appearing in EDE models to date (see e.g. \cite{Poulin:2018cxd, Smith:2020rxx, McDonough:2021pdg}).

It is worth comparing these bounds to bounds on natural inflation from the axion weak gravity, as discussed in e.g. \cite{ArkaniHamed:2006dz, Rudelius:2014wla, delaFuente:2014aca, Rudelius:2015xta, Montero:2015ofa, Brown:2015iha, Heidenreich:2015wga}. In natural inflation, the axion mass will be relatively large, of order $m_\chi \sim 10^{-5} \Mpl$. As a result, the instanton action $S$ can be rather small. Setting $S \gtrsim 1$ for perturbative control of the instanton expansion, the axion weak gravity conjecture implies the familiar constraint $\feff \lesssim \Mpl$, which is indeed satisfied by axions in string theory \cite{Banks:2003sx}. In the case of EDE, however, the axion must be very light, which implies $S \gg 1$ in the absence of significant fine-tuning of the UV scale $\LUV$. This in turn leads to a tighter bound on the axion decay constant, $\feff \ll \Mpl$.

\section{Constraints on Model Parameters}

\begin{center}
\begin{figure*}[t!]
\includegraphics[width=60mm]{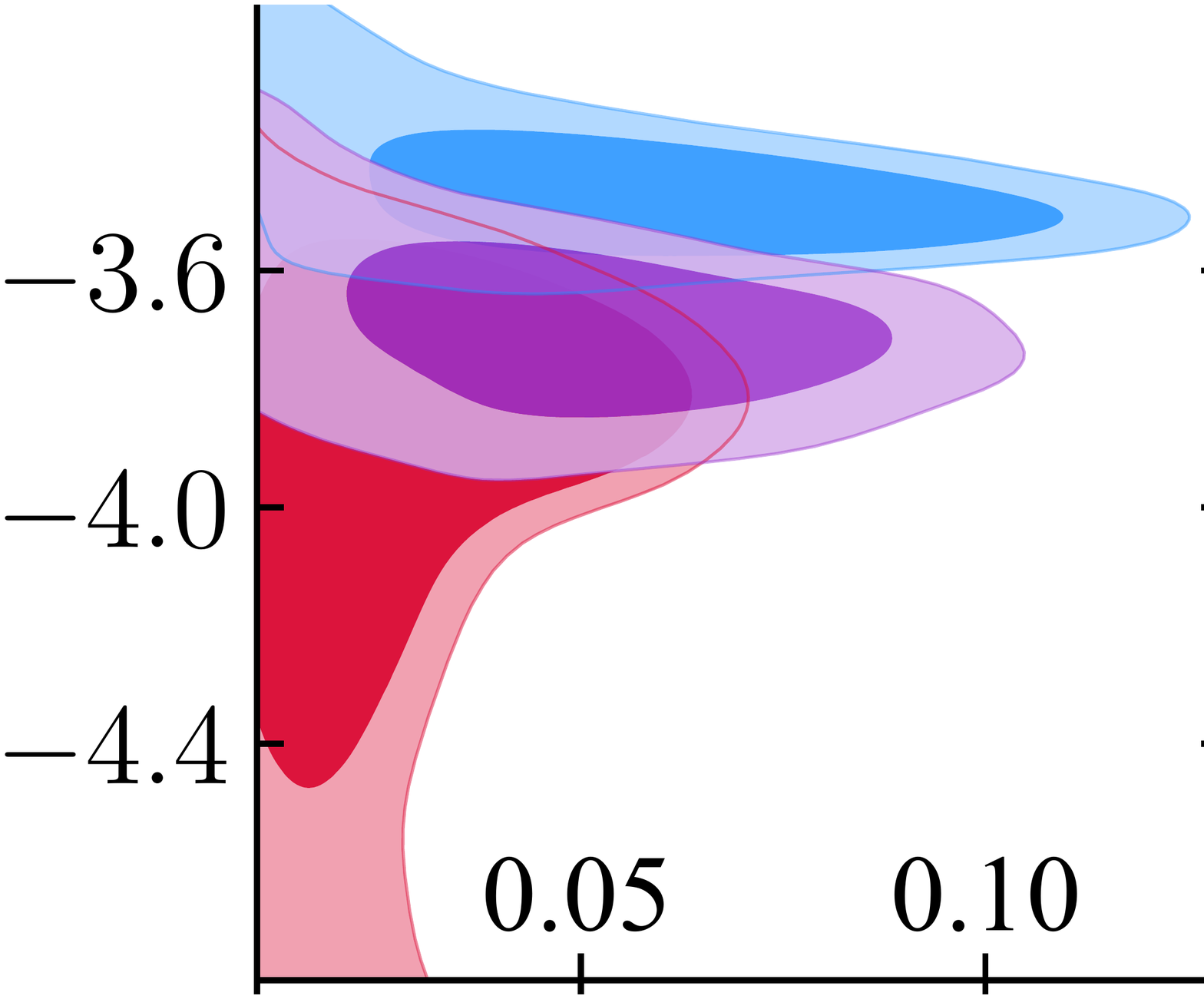} ~~~~
\raisebox{0.25\height}{\includegraphics[width=100mm]{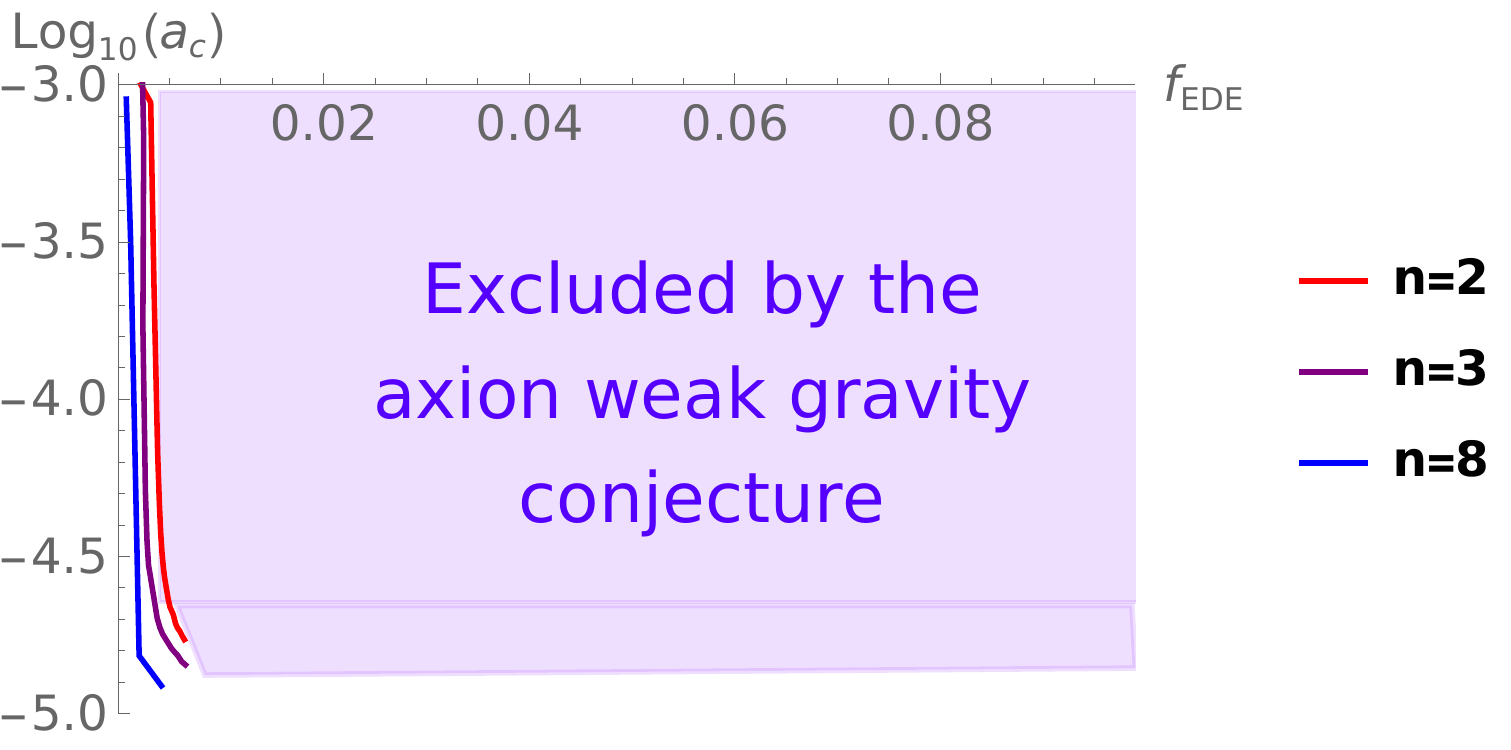}}
\caption{Constraints from the axion weak gravity conjecture for EDE models. The above plots depict the same region of parameter space, with constraints coming from (left) observation and (right) the axion weak gravity conjecture. The left plot (adapted from \cite{Poulin:2018cxd}) depicts observational constraints on $f_{\text{EDE}}$ and $\log_{10}(a_c)$ for $n=2$ (red), $n=3$ (purple) and $n=\infty$ (blue)). The right plot shows the $f=0.008 \Mpl$ curve in $f_{\text{EDE}}$ vs. $\log_{10}(a_c)$ parameter space for $n=2$ (red), $n=3$ (purple), and $n= 8$ (blue). The axion weak gravity conjecture rules out the shaded region to the right of the curves (modulo possible loopholes discussed below).}
\label{fig}
\end{figure*}
\end{center}

So far, we have seen that the axion weak gravity conjecture places constraints on the axion decay constant $f$ in terms of the axion mass parameter $m_\chi$ and the energy scale $\LUV$. In cosmological studies of early dark energy, however, the relevant parameters are instead the critical redshift $z_c \equiv  a_c^{-1} -1$, before which the axion is frozen near the maximum of its potential and effectively acts as a cosmological constant, and the fraction $\fEDE$ of the total energy density stored in the axion condensate at the time $z_c$, $\fEDE \equiv \Omega_\chi(a_c) / \Omega_{\textrm{tot}}(a_c)$.

In this section, we translate the bound $f > 0.008 \Mpl$ derived in the previous section into the space of parameters $a_c$, $\fEDE$ and compare it to the observational constraints on these parameters found in \cite{Poulin:2018cxd}. We do this by taking the standard $\Lambda$CDM model of cosmology, with cosmological parameters
\begin{equation}
\Omega_{0,M} = 0.31\,,~~~~\Omega_{0,R} = 9 \times 10^{-5}\,,~~~~\Omega_\Lambda = 0.69\,,
\end{equation}
and adding an axion $\chi$ with potential \eqref{EDEpotential}, which begins at rest near the top of the cosine hill with an initial position of $\chi_0 = (\pi - 0.01)f$. Following \cite{Poulin:2018cxd}, we then approximate the axion energy density with an EDE energy density:
\begin{equation}
\Omega_\chi(a) = \frac{2  \Omega_\chi(a_c) }{ (a/a_c)^{3(1+ w_n) } + 1}\,,
\end{equation}
where $w_n = (n-1)/(n+1)$. Using a best-fit analysis, we determine the parameters $a_c$, $\fEDE = \Omega_\chi(a_c) / \Omega_{\textrm{tot}}(a_c)$ of the EDE model that most closely approximates the energy density of the axion field condensate. The results of this fitting procedure for a model with $n=3$ are shown in figure \ref{otherfig}.

In general, the EDE parameter $\fEDE$ is approximately half the axion decay constant $f$ (in Planck units) and depends weakly on the axion mass parameter $m_\chi$, whereas the critical redshift $z_c$ is more sensitive to changes in $m_\chi$. The bound $f \lesssim 0.008 \Mpl$ thus roughly translates into a bound $\fEDE \lesssim 0.004$ for typical values of $m_\chi$ for $n=2, 3$, and this bound gets tighter as $n$ increases. More precise bounds are shown in figure \ref{fig} (right), which can be compared with the experimental bounds on $\fEDE$ and $a_c$ from \cite{Poulin:2018cxd}. A large portion of the parameter space is inconsistent with the axion weak gravity conjecture.

Aside from the question of a UV completion, another prominent issue facing EDE is its incompatibility with LSS data; namely, EDE seems to exacerbate the $S_8$ tension \cite{Hill:2020osr, Ivanov:2020ril}. In  \cite{Smith:2020rxx}, it was argued that this tension could be alleviated by taking $\fEDE$ sufficiently large, $\fEDE \gtrsim 0.09$. Here, however, we see that such large values of $\fEDE$ are incompatible with the axion weak gravity conjecture, so consistency between EDE and LSS data likely requires a further late-time alteration of $\Lambda$CDM \cite{Clark:2021hlo}.

\section{Loopholes}

Above, we found a tight constraint on axion decay constants in EDE models. We now ask: can this constraint be loosened?

To begin, it is worth noting that this bound depends only logarithmically on the mass $m_\chi$ and the UV scale $\LUV$, so it is quite robust to changes of these parameters. Decreasing $\LUV$ by a factor of $10^{-10}$, for instance, only loosens the bound \eqref{feffbound} to $\feff \lesssim 0.13 \Mpl$.

Rather than changing the parameters of the model, another way to loosen the constraint would be to change the parameter $c$ appearing in the axion weak gravity conjecture. The maximum value of the axion decay constant scales linearly with $c$, so if $c$ were increased by a factor of 3, the maximal value of $\feff$ would also grow by a factor of 3, accommodating a wider range of parameter space. However, at present, there is no competing proposal for $c$ in theories without massless scalars other than the one introduced in \eqref{ceq}, so there is no compelling justification for loosening the constraint in this way. Clearly, the relevance to EDE models makes the task of determining the correct value of $c$ even more urgent.

Constraints from the axion weak gravity conjecture on natural inflation models feature a pair of famous loopholes, which ostensibly could be used as loopholes for EDE models as well. Most prominent is the \emph{extra instanton loophole} \cite{delaFuente:2014aca, Rudelius:2015xta, Brown:2015iha, Hebecker:2015rya}, in which one instanton satisfies the axion weak gravity conjecture, while another gives the dominant contribution to the axion potential. Since the latter instanton controls the phenomenology of the model, the constraints from the axion weak gravity conjecture are rendered vacuous.

To be more explicit, let us suppose that the dominant contribution to the axion potential comes from an instanton of charge 1, with action $S_1$. Let us suppose that the axion weak gravity conjecture is satisfied by an instanton of charge $k$, with action $S_k$. Thus, we have
\begin{equation}
\frac{f S_{k}  }{k} \leq  \frac{\sqrt{6} \pi }{4} \Mpl\,.
\end{equation}
In order to suppress the contributions from the charge $k$ instanton relative to those from the dominant charge $1$ instanton, we need $S_k > S_1$, which therefore implies
\begin{equation}
\frac{f S_{1}  }{k} \leq  \frac{\sqrt{6} \pi }{4} \Mpl\,.
\end{equation}
This weakens the bound \eqref{generalbound} by a factor of $k$, so by taking $k$ sufficiently large, the axion weak gravity conjecture bounds on $S_1$ become negligible. However, in practice, every string compactification studied thus far satisfies the weak gravity conjecture with a state of charge $k \leq 3$ \cite{Heidenreich:2016aqi, Lee:2019tst}, so parametric enhancement of $k$ seems difficult to achieve in a UV complete theory. Furthermore, this loophole introduces additional parameters into the theory (namely, the instanton actions $S_2$, ..., $S_k$) which must be tuned appropriately to ensure consistency with the axion weak gravity conjecture without spoiling the phenomenology of the model. Such tuning lowers the prior probability of the model in a Bayesian framework.

The other prominent loophole in the natural inflation literature is known as the \emph{small action loophole}. In some cases, such as extranatural inflation \cite{Arkani-Hamed:2003xts}, it may be possible to suppress higher harmonics to the axion potential even for $S < 1$, which in principle could lead to a parametrically large decay constant \cite{delaFuente:2014aca}. In the case of EDE, however, $S \gg 1$ is required not only for perturbative control of the instanton expansion, but also for the lightness of the axion field, $m_\chi \ll$ eV. The small action loophole is therefore unavailable in the EDE context.

Finally, let us remark on the possibility evading the constraint \eqref{generalbound} using multi-axion models of EDE. Historically, multi-axion models of natural inflation \cite{liddle:1998jc, Kim:2004rp, dimopoulos:2005ac} have been a popular approach to evading axion decay constant constraints from string theory \cite{Banks:2003sx} and the axion weak gravity conjecture \cite{ArkaniHamed:2006dz}. However, more recent studies have shown that the axion weak gravity conjecture constraints extend straightforwardly to multi-axion models \cite{Rudelius:2015xta, Montero:2015ofa, Brown:2015iha, Heidenreich:2015wga}, so these models offer no additional path to EDE aside from the loopholes discussed above.

We conclude that constraints on EDE from the axion weak gravity conjecture are quite robust. Loopholes do exist, but it may well prove more difficult to thread such a loophole than it is to build an EDE model satisfying the axion weak gravity conjecture constraint \eqref{generalbound}.


\section{Discussion}\label{sec:DISC}

We have seen that the requirement of a consistent UV completion places significant constraints on models of early dark energy. To fully understand how significant these constraints are, however, several questions need to be answered.

First of all, is it possible to generate an EDE model with $n \geq 2$ through a delicate cancelation of terms in an instanton expansion? How delicate does this cancelation have to be to produce a phenomenologically viable model of EDE? In other words, how fine-tuned would the mass of an EDE axion have to be to resolve the Hubble tension?

Is $c= \sqrt{6} \pi/4$ the correct value to use in the axion weak gravity conjecture? If not, what is the correct value of $c$? If the answer is much larger (smaller) than this value, the constraint on EDE models will be much weaker (stronger) than the one proposed in this paper.

From a model-building perspective, is there a way to thread the extra instanton loophole detailed above, in which one instanton satisfies the axion weak gravity conjecture while another produces a model of early dark energy? Is there a way to tune the scale $\LUV$ by many orders of magnitude, thereby loosening the bounds on the axion decay constant $f$? Can phenomenologically viable models satisfy the bound $f < 0.008 \Mpl$? Is there a phenomenologically viable axio-dilaton model of EDE with $n=1$ and $f < 0.008 \Mpl$?

Besides the axion weak gravity conjecture, do other swampland conjectures give significant constraints on EDE models?  In the same paper that first introduced the swampland distance conjecture \cite{Ooguri:2006in}, Ooguri and Vafa also argued that scalar moduli space should have no non-trivial 1-cycles with minimum length in a given homotopy class. This means that any periodic axion field $\chi$ must be accompanied by a dilaton $\phi$, such that the decay constant of $\chi$ depends on $\phi$, $f=f(\phi)$, with $f \rightarrow 0$ as $\phi \rightarrow \infty$. The swampland distance conjecture constrains the limit $\phi \rightarrow \infty$, implying a tower of light states with masses beginning at the scale $e^{- \alpha \phi}$, where $\alpha$ is $O(\Mpl)$. It would be worthwhile to study how introducing dilaton-dependence to the axion decay constant, $f=f(\phi)$, affects the analysis of \cite{Alexander:2019rsc}. It may also prove interesting to study the cosmological consequences of an $e^{- \alpha \phi}\bar \psi \psi$ dilatonic coupling to dark matter in such an axio-dilaton EDE model.\footnote{One recent paper \cite{McDonough:2021pdg} considered the cosmological consequences of a coupling $e^{-\alpha \chi} \bar \psi \psi$ between a fermionic dark matter field $\psi$ and the EDE axion $\chi$. Such couplings have been observed in string compactifications \cite{Baume:2016psm} and are a prediction of the \emph{refined} swampland distance conjecture \cite{Klaewer:2016kiy}, but they are not necessarily required by the original swampland distance conjecture \cite{Ooguri:2006in} because the periodicity of the axion under $\chi \rightarrow \chi + 2 \pi f$ means that $\chi \rightarrow \infty$ is not an infinite distance limit in moduli space.}

Of course, these questions are particularly important insofar as the Hubble tension remains unresolved. The most important question---will the Hubble tension stand up to increased experimental scrutiny?---remains to be seen.

\vspace*{1cm}
{\bf Acknowledgments.}  We thank Joel Primack and Matthew Reece for useful discussions. We thank Vivian Poulin for permission to reprint figure \ref{fig} (left) from \cite{Poulin:2018cxd}. We gratefully acknowledge the hospitality of the University of California, Santa Cruz, where a portion of this work was completed. This work was supported in part by the Berkeley Center for Theoretical Physics; by the Department of Energy, Office of Science, Office of High Energy Physics under QuantISED Award DE-SC0019380 and under contract DE-AC02-05CH11231; and by the National Science Foundation under Award Number 2112880.


\bibliography{ref}

\end{document}